\documentclass[preprintnumbers,amsmath,amssymbm,prd]{revtex4}
\usepackage{epsfig}
\usepackage{graphicx}

\begin{document}
\title{Non-equatorial scalar rings supported by magnetized Schwarzschild-Melvin black holes}
\author{Shahar Hod}
\affiliation{The Ruppin Academic Center, Emeq Hefer 40250, Israel}
\affiliation{ }
\affiliation{The Hadassah Institute, Jerusalem 91010, Israel}
\date{\today}

\begin{abstract}
\ \ \ It has recently been demonstrated that magnetized black holes in composed Einstein-Maxwell-scalar-Gauss-Bonnet field 
theories with a non-minimal negative coupling of the scalar field to the Gauss-Bonnet curvature 
invariant may support spatially regular scalar hairy configurations. In particular, it has been revealed that, 
for Schwarzschild-Melvin black-hole spacetimes, the onset of the near-horizon spontaneous scalarization phenomenon is marked by the 
numerically computed dimensionless critical relation $(BM)_{\text{crit}}\simeq0.971$, 
where $\{M,B\}$ are respectively the mass and the magnetic field of the spacetime. 
In the present paper we prove, using analytical techniques, that the boundary 
between bald Schwarzschild-Melvin black-hole spacetimes and 
hairy (scalarized) black-hole solutions of the composed Einstein-Maxwell-scalar-Gauss-Bonnet theory is characterized by the 
exact dimensionless relation $(BM)_{\text{crit}}=\sqrt{{{\sqrt{6}-2}\over{2\sqrt{6}}}+\sqrt{{{\sqrt{6}-1}\over{2}}}}$ for 
the critical magnetic strength. 
Intriguingly, we prove that the critical dimensionless magnetic parameter $(BM)_{\text{crit}}$ 
corresponds to magnetized black holes that support a pair of linearized non-minimally coupled thin scalar rings that 
are characterized by the non-equatorial polar angular relation 
$(\sin^2\theta)_{\text{scalar-ring}}={{690-72\sqrt{6}+4\sqrt{3258\sqrt{6}-7158}}\over{789}}<1$. 
It is also proved
that the classically allowed angular region for the negative-coupling near-horizon spontaneous scalarization phenomenon
of magnetized Schwarzschild-Melvin spacetimes is restricted to the black-hole poles, $\sin^2\theta_{\text{scalar}}\to0$, 
in the asymptotic large-strength magnetic regime $BM\gg1$.
\end{abstract}
\bigskip
\maketitle

\section{Introduction}

The celebrated no-hair conjecture in black-hole physics \cite{NHC,JDB} has asserted that static scalar field configurations 
cannot be supported in black-hole spacetimes that contain spatially regular absorbing horizons. 
Early mathematical investigations of the 
Einstein-matter field equations \cite{Bek1,Sot2,Her1,Sot3,BekMay,Hod1} have revealed, in accord with the spirit 
of this influential conjecture, that black holes cannot support scalar fields which are minimally coupled to the Ricci scalar of 
the curved spacetime. Similar conclusions have been obtained for scalar fields with a non-trivial (non-minimal) coupling 
to the Ricci curvature scalar \cite{BekMay,Hod1}.  

However, recent mathematical studies \cite{Sot5,Sot1,GB1,GB2,ChunHer,SotN,Hodsg1,Hodsg2,Hodca,Done,Herrecnum,BeCo,Brih,Hernn} 
of the coupled Einstein-matter field equations have revealed the physically intriguing fact that black holes 
with spatially regular horizons can support hairy matter configurations which are made of scalar fields with a 
direct non-minimal coupling to the 
Gauss-Bonnet invariant ${\cal G}$ of the curved black-hole spacetime \cite{Notechar,Hersc1,Hersc2,Hodsc1,Hodsc2,Moh,Hodnrx}. 

The critical boundary between bald black holes and hairy (scalarized) black-hole spacetimes in Einstein-Gauss-Bonnet field theories 
is marked by the presence of `cloudy' configurations, linearized non-minimally coupled scalar fields which 
are supported by the familiar black-hole solutions of the Einstein field equations. In particular, depending on the matter content of the theory and the assumed symmetry of the spacetime, cloudy non-minimally coupled scalar field configurations have 
been studied in Schwarzschild, Reissner-Nordstr\"om, Kerr, and Kerr-Newman black-hole spacetimes \cite{Sot5,Sot1,GB1,GB2,ChunHer,SotN,Hodsg1,Hodsg2,Hodca,Done,Herrecnum,BeCo,Brih,Hernn}. 

The spontaneous scalarization phenomenon of black holes in Einstein-Gauss-Bonnet field theories is closely related to the fact that the Klein-Gordon wave equation of the non-minimally coupled scalar field $\phi$ 
contains an effective spatially-dependent mass term. This mass term, which has the compact linearized 
form $-\eta\phi{\cal G}$ [see Eq. (\ref{Eq10}) below] \cite{Noteetaa}, 
may become {\it negative} in the vicinity of the black-hole horizon, implying that the effective potential of the composed 
black-hole-scalar-field system may behave as an attractive potential well that binds the scalar field 
to the near-horizon region of the central black hole.

Intriguingly, it has recently been demonstrated in the physically important work \cite{Hernw} that, 
in Einstein-Gauss-Bonnet field theories, the phenomenon of black-hole spontaneous scalarization 
can be triggered by the presence of magnetic fields. 
In particular, it has been demonstrated numerically in \cite{Hernw} that, 
for Schwarzschild-Melvin black-hole spacetimes, the spontaneous scalarization phenomenon is 
magnetically-induced in the sense that only black holes in the dimensionless strong magnetic regime  
\begin{equation}\label{Eq1}
BM>(BM)_{\text{crit}}\simeq0.971\ 
\end{equation}
can support scalar field configurations with a non-minimal {\it negative} coupling to the Gauss-Bonnet invariant 
of the curved magnetized spacetime 
(Here $M$ and $B$ are respectively the mass and the magnetic field of the black-hole spacetime). 

The main goal of the present compact paper is to explore, using analytical techniques, the onset of the spontaneous scalarization phenomenon in magnetized Schwarzschild-Melvin black-hole spacetimes of the 
composed Einstein-Maxwell-scalar-Gauss-Bonnet field theory with negative values of the non-minimal 
coupling parameter $\eta$. 

In particular, below we shall derive a remarkably compact analytical expression for the
critical magnetic strength $(BM)_{\text{crit}}$ which marks the boundary between scalarless Schwarzschild-Melvin black holes and 
scalarized (hairy) magnetized black-hole configurations. 
In addition, we shall reveal the physically intriguing
fact that the black-hole solutions of the composed Einstein-Maxwell-scalar-Gauss-Bonnet field theory can 
support pairs of infinitesimally thin static scalar rings which are located 
above and below the equator of the central supporting black hole. 

\section{Description of the system}

We shall study, using analytical techniques, the onset of the negative-coupling spontaneous scalarization 
phenomenon of magnetized Schwarzschild-Melvin black holes in the composed Einstein-Maxwell-scalar-Gauss-Bonnet field 
theory whose action is given by the integral expression \cite{Hernw}
\begin{equation}\label{Eq2}
S={{1}\over{16\pi}}\int
d^4x\sqrt{-g}\Big[R-F_{\mu\nu}F^{\mu\nu}-{1\over2}\nabla_{\alpha}\phi\nabla^{\alpha}\phi
+f(\phi){\cal G}\Big]\  ,
\end{equation}
where $F_{\mu\nu}$ is the Maxwell electromagnetic tensor of the spacetime. 
The term $f(\phi){\cal G}$, which is essential for the existence of spontaneously scalarized black-hole solutions, 
represents the direct (non-minimal) coupling between the massless scalar field $\phi$ and the Gauss-Bonnet 
invariant 
\begin{equation}\label{Eq3}
{\cal G}\equiv R^2-4R_{\mu\nu}R^{\mu\nu}+R_{\mu\nu\alpha\beta}R^{\mu\nu\alpha\beta}\
\end{equation}
of the black-hole spacetime. 

The magnetized Schwarzschild-Melvin black-hole solution of the composed field theory (\ref{Eq2}) is 
characterized by the curved line element \cite{Hernw,Melv,HF,Notecord,Noteunits}
\begin{equation}\label{Eq4}
ds^2=\Lambda^2[-f(r)dt^2+f^{-1}(r)dr^2+r^2d\theta^2]+\Lambda^{-2}r^2\sin^2\theta d\varphi^2\  ,
\end{equation}
where 
\begin{equation}\label{Eq5}
\Lambda=\Lambda(r,\theta;B)=1+{1\over4}B^2r^2\sin^2\theta\  . 
\end{equation}
The metric function
\begin{equation}\label{Eq6}
f(r)=1-{{2M}\over{r}}\  ,
\end{equation}
whose simple root
\begin{equation}\label{Eq7}
r_{\text{H}}=2M\  
\end{equation}
determines the radius of the black-hole horizon, 
is expressed in terms of the mass $M$ of the spacetime. 

From Eqs. (\ref{Eq3}), (\ref{Eq4}), (\ref{Eq5}), and (\ref{Eq6}) one obtains 
the (rather cumbersome) near-horizon dimensionless functional expression \cite{Hernw,NoteMev}
\begin{eqnarray}\label{Eq8}
M^4{\cal G}(r\to r_{\text{H}}^+,\theta;M,B)=\Big[4\big(1+B^2M^2\sin^2\theta\big)^8\Big]^{-1}\times
\Big\{3\Big[B^8M^8\sin^8\theta-2B^4M^4\sin^4\theta+\big[1-B^4M^4\sin^2(2\theta)\big]^2\Big]+\nonumber \\
24\cos^2\theta\big(B^8M^8\sin^6\theta-B^6M^6\sin^4\theta+B^2M^2\big)+
16B^4M^4\cos^4\theta\big(1-6B^2M^2\sin^2\theta\big)\Big\}\
\end{eqnarray}
for the magnetically-dependent Gauss-Bonnet invariant of the Schwarzschild-Melvin black hole. 

The action (\ref{Eq2}) of the composed Einstein-Maxwell-scalar-Gauss-Bonnet field theory 
yields the characteristic Klein-Gordon differential equation \cite{Hernw}
\begin{equation}\label{Eq9}
\nabla^\nu\nabla_{\nu}\phi=\mu^2_{\text{eff}}\phi\
\end{equation}
for the scalar field, where the presence of the magnetically-dependent effective mass term in (\ref{Eq9}),
\begin{equation}\label{Eq10}
\mu^2_{\text{eff}}(\theta;M,B)=-\eta\cdot{\cal G}(\theta;M,B)\  ,
\end{equation}
is a direct outcome of the non-minimal coupling between the scalar field of the theory and the 
Gauss-Bonnet invariant of the spacetime. 
The physical parameter $\eta$ in (\ref{Eq10}) plays the role of an expansion coefficient 
in the weak-field functional behavior \cite{Hernw}
\begin{equation}\label{Eq11}
f(\phi)={1\over2}\eta\phi^2\ 
\end{equation}
of the scalar coupling function. 
This physical parameter \cite{Noteett} determines the strength of the direct non-minimal coupling 
between the massless scalar field and the Gauss-Bonnet invariant (\ref{Eq8}) 
of the magnetized Schwarzschild-Melvin black hole. 

\section{Onset of negative-coupling near-horizon spontaneous scalarization in magnetized Schwarzschild-Melvin black-hole spacetimes}

In the present section we shall study the onset of the negative-coupling 
spontaneous scalarization phenomenon in the composed Einstein-Maxwell-scalar-Gauss-Bonnet field 
theory (\ref{Eq2}). 
In particular, we shall reveal the interesting fact that, at the onset of the magnetically-induced 
spontaneous scalarization phenomenon, Schwarzschild-Melvin black-hole spacetimes can support 
non-equatorial thin rings which are made of massless scalar fields with a negative coupling to the Gauss-Bonnet invariant 
of the magnetized spacetime. 

The presence of an effective {\it negative} mass term (a binding potential well) 
in the Klein-Gordon wave equation (\ref{Eq9}) of the non-minimally coupled 
scalar field provides a necessary condition for the existence of spatially regular 
supported field configurations (bound-state scalar clouds) 
in the black-hole spacetime \cite{GB1,Done,ChunHer,Hodca,Hernn}. 
Intriguingly, from Eq. (\ref{Eq8}) one learns that, depending on the values of the 
dimensionless magnetic parameter $BM$ of the spacetime and the polar angle $\theta$, 
the effective mass term (\ref{Eq10}) in the Klein-Gordon differential equation (\ref{Eq9}) 
may become negative in the vicinity of the horizon of the 
Schwarzschild-Melvin black hole. 

In particular, the {\it onset} of the near-horizon spontaneous scalarization phenomenon 
in the magnetized black-hole spacetime (\ref{Eq4}) 
is marked by the critical functional relation \cite{Hodca,Hernn,Hodjp}
\begin{equation}\label{Eq12}
\text{min}\{\mu^2_{\text{eff}}(\theta;M,B)\}\to 0^{-}\  .
\end{equation}
For negative values of the non-minimal coupling parameter $\eta$ of the 
composed Einstein-Maxwell-scalar-Gauss-Bonnet field theory (\ref{Eq2}), the characteristic 
relation (\ref{Eq12}) yields the critical functional relation
\begin{equation}\label{Eq13}
\text{min}\{G(\theta;M,B)\}\to 0^{-}\
\end{equation}
at the onset of the spontaneous scalarization phenomenon. 

Interestingly, and most importantly for our analysis, as we shall now show explicitly, 
the set of coupled equations [see Eq. (\ref{Eq13})]
\begin{equation}\label{Eq14}
G(\theta;M,B)=0\
\end{equation}
with 
\begin{equation}\label{Eq15}
{{dG(\theta;M,B)}\over{d\theta}}=0\  ,
\end{equation}
which determine the onset of the near-horizon spontaneous scalarization phenomenon of the 
magnetized black-hole spacetime (\ref{Eq4}), can be solved {\it analytically}. 

To see this, it is convenient to define the dimensionless variables
\begin{equation}\label{Eq16}
\beta\equiv (BM)^2\ \ \ \ ; \ \ \ \ x\equiv\sin^2\theta\  ,
\end{equation}
in terms of which the near-horizon Gauss-Bonnet invariant (\ref{Eq8}) can be written 
in the dimensionless form \cite{Notesim}
\begin{eqnarray}\label{Eq17}
M^4{\cal G}(r\to r_{\text{H}}^+,x;\beta)=\big[4(1+\beta x)^8\big]^{-1}\times
\Big\{3\Big[\beta^4x^4-2\beta^2x^2+\big[1-4\beta^2x(1-x)\big]^2\Big]+\nonumber \\
24(1-x)\big(\beta^4x^3-\beta^3x^2+\beta\big)+
16\beta^2(1-x)^2\big(1-6\beta x\big)\Big\}  .\
\end{eqnarray}
Substituting (\ref{Eq17}) into Eqs. (\ref{Eq14}) and (\ref{Eq15}), one obtains the coupled equations
\begin{equation}\label{Eq18}
(16\beta^2+24\beta+3)-8\beta(12\beta^2+7\beta+3)x+2\beta^2 (24\beta^2+84\beta+17)x^2-72\beta^3(\beta+1)x^3
+27\beta^4x^4=0\ 
\end{equation}
and \cite{Notedu}
\begin{equation}\label{Eq19}
-2(12\beta^2+7\beta+3)+\beta(24\beta^2+84\beta+17)x-54\beta^2(\beta+1)x^2+27\beta^3x^3=0\  
\end{equation}
for the critical dimensionless variables $\{x_{\text{crit}},\beta_{\text{crit}}\}$.

Interestingly, the coupled polynomial equations (\ref{Eq18}) and (\ref{Eq19}), which determine the onset of the near-horizon
spontaneous scalarization phenomenon of the magnetized Schwarzschild-Melvin black hole (\ref{Eq4}), can be solved 
analytically to yield the remarkably compact critical solution
\begin{equation}\label{Eq20}
\beta_{\text{crit}}={1\over2}-{{1}\over{\sqrt{6}}}+\sqrt{{{\sqrt{6}-1}\over{2}}}\
\end{equation}
with
\begin{equation}\label{Eq21}
x_{\text{crit}}={{690-72\sqrt{6}+4\sqrt{3258\sqrt{6}-7158}}\over{789}}\  .
\end{equation}
From the analytically derived expression (\ref{Eq21}) one deduces that the onset of the spontaneous 
scalarization phenomenon at the critical magnetic strength (\ref{Eq20}) is characterized 
by the presence of a pair of infinitesimally thin non-equatorial scalar rings which are supported by the magnetized 
black hole at the unique polar angles $\theta^{-}_{\text{ring}}=63.177^{\circ}$ 
and $\theta^{+}_{\text{ring}}=116.823^{\circ}$ [see Eqs. (\ref{Eq16}) and (\ref{Eq21})] \cite{Notering}.

\section{The classically allowed polar angular region for the spontaneous scalarization phenomenon of 
magnetized Schwarzschild-Melvin black holes}

In the present section we shall reveal the physically interesting fact that the negative-coupling near-horizon 
spontaneous scalarization phenomenon of Schwarzschild-Melvin black holes is characterized by two critical values of the
dimensionless magnetic strength parameter $BM$. 
These critical magnetic strengths mark the boundaries between 
three qualitatively different spatial behaviors of the 
composed magnetized-black-hole-nonminimally-coupled-scalar-field configurations.  

We shall now discuss the three qualitatively different regimes of the magnetic parameter $BM$:

(1) Case I: In the dimensionless weak magnetic regime
\begin{equation}\label{Eq22}
\beta<\beta^{-}_{\text{crit}}\equiv{1\over2}-{{1}\over{\sqrt{6}}}+\sqrt{{{\sqrt{6}-1}\over{2}}}\  ,
\end{equation}
the near-horizon Gauss-Bonnet invariant (\ref{Eq8}) of the Schwarzschild-Melvin black hole is positive definite 
in the entire polar angular range $\theta\in[0,\pi]$. 
Thus, Schwarzschild-Melvin black holes in the dimensionless magnetic regime (\ref{Eq22}) cannot support 
negatively coupled massless scalar fields in their near-horizon region.

(2) Case II: In the intermediate-strength magnetic regime
\begin{equation}\label{Eq23}
{1\over2}-{{1}\over{\sqrt{6}}}+\sqrt{{{\sqrt{6}-1}\over{2}}}\equiv\beta^{-}_{\text{crit}}\leq\beta
\leq\beta^{+}_{\text{crit}}\equiv1\  ,
\end{equation}
the near-horizon Gauss-Bonnet invariant (\ref{Eq8}) of the Schwarzschild-Melvin black hole becomes negative 
in the polar angular range 
\begin{eqnarray}\label{Eq24}
{1\over3}\Bigg[2+{{2-2\sqrt{{{2}\over{3}}}}\over{\beta}}-{{\sqrt{12-{{12-4\sqrt{6}}\over{\beta}}-{{8\sqrt{6}-11}
\over{\beta^2}}}}\over{\sqrt{3}}}\Bigg]\leq\sin^2\theta\leq
{1\over3}\Bigg[2+{{2-2\sqrt{{{2}\over{3}}}}\over{\beta}}+{{\sqrt{12-{{12-4\sqrt{6}}\over{\beta}}-{{8\sqrt{6}-11}
\over{\beta^2}}}}\over{\sqrt{3}}}\Bigg]\  .
\end{eqnarray}
The polar range (\ref{Eq24}) defines, in the dimensionless magnetic regime (\ref{Eq23}), 
the classically allowed angular region for the spontaneous scalarization phenomenon of negatively-coupled scalar 
fields in the near-horizon region of the magnetized Schwarzschild-Melvin black hole. 

Interestingly, from the analytically derived relation (\ref{Eq24}) one learns that the 
classically allowed angular width for the near-horizon spontaneous scalarization phenomenon is infinitesimally thin in 
the near-critical magnetic regime $\beta/\beta^{-}_{\text{crit}}\to1^+$. In particular, defining the 
near-critical relation
\begin{equation}\label{Eq25}
\beta=\beta^{-}_{\text{crit}}\cdot(1+\epsilon)\ \ \ , \ \ \ 0\leq\epsilon\ll1\  ,
\end{equation}
one finds from (\ref{Eq24}) the non-trivial (non-linear) critical functional behavior 
\begin{equation}\label{Eq26}
\Delta(\sin^2\theta)={{\sqrt{{32\alpha\over3}\Big(3\sqrt{2}-2\sqrt{3}+6\alpha\Big)}}\over{2+3\sqrt{2}\alpha-\alpha^2}}\cdot\sqrt{\epsilon}\ \ \ , \ \ \ \alpha\equiv\sqrt{\sqrt{6}-1}\
\end{equation}
for the infinitesimally thin classically allowed angular interval. 

The classically allowed angular region (\ref{Eq24}) for the near-horizon spontaneous scalarization phenomenon of the 
magnetized Schwarzschild-Melvin black holes in the intermediate magnetic regime (\ref{Eq23}) 
is a monotonically increasing function of the magnetic strength parameter $\beta$. 
In particular, it is characterized by the limiting property
\begin{equation}\label{Eq27}
{{15-4\sqrt{6}}\over{9}}\leq\sin^2\theta\leq1\ \ \ \ \text{for}\ \ \ \ \beta=\beta^{+}_{\text{crit}}\  
\end{equation}
for the critical magnetic strength $\beta=\beta^{+}_{\text{crit}}$. 

It is interesting to point out that the black-hole configuration with the critical magnetic strength $\beta^{+}_{\text{crit}}=1$ 
is unique in the sense that 
the classically allowed polar angular region for the negative-coupling near-horizon spontaneous scalarization phenomenon 
extends in this case all the way to the equator of the magnetized black hole. 

(3) Case III: In the dimensionless strong magnetic regime
\begin{equation}\label{Eq28}
1\equiv\beta^{+}_{\text{crit}}<\beta\  ,
\end{equation}
the near-horizon Gauss-Bonnet invariant (\ref{Eq8}) of the Schwarzschild-Melvin black holes becomes negative 
in the polar angular range 
\begin{eqnarray}\label{Eq29}
{1\over3}\Bigg[2+{{2-2\sqrt{{{2}\over{3}}}}\over{\beta}}-{{\sqrt{12-{{12-4\sqrt{6}}\over{\beta}}-{{8\sqrt{6}-11}
\over{\beta^2}}}}\over{\sqrt{3}}}\Bigg]\leq\sin^2\theta\leq
{1\over3}\Bigg[2+{{2+2\sqrt{{{2}\over{3}}}}\over{\beta}}-{{\sqrt{12-{{12+4\sqrt{6}}\over{\beta}}+{{8\sqrt{6}+11}
\over{\beta^2}}}}\over{\sqrt{3}}}\Bigg]\  .
\end{eqnarray}
The polar range (\ref{Eq29}) defines the classically allowed angular region for the negative-coupling near-horizon spontaneous 
scalarization phenomenon of strongly magnetized [see Eq. (\ref{Eq28})] Schwarzschild-Melvin black holes.

Interestingly, one finds that, in the strong magnetic regime (\ref{Eq28}), the two polar boundaries of the classically allowed region (\ref{Eq29}) are monotonically decreasing functions of the dimensionless 
magnetic parameter $\beta$. 
In particular, one finds that the classically allowed angular interval (\ref{Eq29}) 
gradually shrinks to the infinitesimally thin polar region
\begin{equation}\label{Eq30}
{{1-\sqrt{{{2}\over{3}}}}\over{\beta}}\leq\sin^2\theta\leq{{1+\sqrt{{{2}\over{3}}}}\over{\beta}}\ \ \ \ \ \text{for}\ \ \ \ \ \beta\to\infty\  
\end{equation}
in the asymptotically large magnetic regime. One therefore concludes that the classically allowed polar region 
for the negative-coupling near-horizon spontaneous scalarization phenomenon of the 
magnetized Schwarzschild-Melvin spacetime is restricted to the vicinity of the black-hole poles in the asymptotic
large-strength magnetic regime $\beta\gg1$. 

\section{Summary and discussion}

It has recently been shown \cite{Hernw} that magnetized black holes may support non-minimally 
coupled scalar hairy configurations. 
In particular, the recently published important work \cite{Hernw} has
revealed the physically interesting fact that, in composed Einstein-Maxwell-scalar-Gauss-Bonnet field 
theories with negative values of the
non-minimal coupling parameter $\eta$, there exists a critical magnetic strength [see Eq. (\ref{Eq1})],
\begin{equation}\label{Eq31}
(BM)_{\text{crit}}\simeq0.971\  ,
\end{equation}
above which Schwarzschild-Melvin black holes can support near-horizon massless scalar field 
configurations with a non-minimal direct coupling to the magnetically-dependent Gauss-Bonnet curvature invariant (\ref{Eq8}). 

In the present paper we have studied, using {\it analytical} techniques, 
the onset of the negative-coupling near-horizon spontaneous scalarization phenomenon in magnetized Schwarzschild-Melvin 
black-hole spacetimes. The main results derived in this paper and their physical implications are as follows:

(1) We have proved that the critical black-hole magnetic strength 
$({BM})_{\text{crit}}$, which marks the boundary between 
bald Schwarzschild-Melvin black holes and hairy (scalarized) black holes in
the Einstein-Maxwell-scalar-Gauss-Bonnet field theory (\ref{Eq2}) with a negative non-minimal Gauss-Bonnet-scalar-field coupling, 
is given by the compact dimensionless analytical expression [see Eqs. (\ref{Eq16}) and (\ref{Eq20})]
\begin{equation}\label{Eq32}
(BM)_{\text{crit}}=\sqrt{{1\over2}-{{1}\over{\sqrt{6}}}+\sqrt{{{\sqrt{6}-1}\over{2}}}}\  .
\end{equation}
It is worth emphasizing the fact that the analytically derived
critical black-hole magnetic strength (\ref{Eq32}) agrees
remarkably well with the corresponding numerical value (\ref{Eq31}) of the critical 
magnetic field as originally presented in the physically interesting work \cite{Hernw}.

(2) We have revealed the fact that {\it non}-equatorial thin scalar rings can be supported in magnetized black-hole spacetimes. 
In particular, one finds that, in the dimensionless critical limit
\begin{equation}\label{Eq33}
BM\to[(BM)_{\text{crit}}]^+\  ,
\end{equation}
the effective mass term (\ref{Eq10}) of the composed 
Schwarzschild-Melvin-black-hole-nonminimally-coupled-massless-scalar-field 
system becomes {\it negative} in a pair of narrow non-equatorial polar rings which are 
characterized by the analytically derived angular relation [see Eqs. (\ref{Eq16}), (\ref{Eq21}), (\ref{Eq25}) and (\ref{Eq26})]
\begin{equation}\label{Eq34}
\sin^2\theta_{\text{ring}}={{18+8(3\sqrt{2}-2\sqrt{3}){\alpha}}\over{\big(2+3\sqrt{2}\alpha-\alpha^2\big)^2}}
\ \ \ , \ \ \ \alpha\equiv\sqrt{\sqrt{6}-1}
\  .
\end{equation}
This physically interesting property of the Schwarzschild-Melvin curved spacetime implies that magnetized black holes 
can support, in the critical limit (\ref{Eq33}), cloudy 
configurations of the non-minimally coupled massless scalar fields in the form of two infinitesimally thin non-equatorial 
scalar rings which are characterized by the polar angles [see Eq. (\ref{Eq34})]
\begin{equation}\label{Eq35}
\theta^{-}_{\text{ring}}=63.177^{\circ}\ \ \ ; \ \ \ \theta^{+}_{\text{ring}}=116.823^{\circ}\  .
\end{equation}

(3) It has been revealed that the Schwarzschild-Melvin black hole with the critical magnetic strength $BM=1$ 
is unique in the sense that the near-horizon angular region for which the Gauss-Bonnet invariant becomes 
non-positive (thus allowing for the onset of the negative-coupling spontaneous scalarization phenomenon) 
extends in this case all the way to the equator of the magnetized black hole [see Eq. (\ref{Eq24})]. 

(4) We have revealed the fact that, for strong magnetic fields in the $BM\gg1$ regime, 
the classically allowed angular region for the negative-coupling spontaneous scalarization phenomenon of magnetized Schwarzschild-Melvin spacetimes gradually shrinks as the value of the dimensionless 
magnetic strength $BM$ increases. 
In particular, we have proved that the near-horizon spontaneous scalarization phenomenon 
is restricted to the narrow near-polar angular interval [see Eqs. (\ref{Eq16}) and (\ref{Eq30})] \cite{Notesymm} 
\begin{equation}\label{Eq36}
{{\sqrt{1-\sqrt{{{2}\over{3}}}}}\over{BM}}\leq\theta_{\text{scalar}}
\leq{\sqrt{{1+\sqrt{{{2}\over{3}}}}}\over{BM}}\ \ \ \ \ \text{for}\ \ \ \ \ BM\gg1\
\end{equation}
in the asymptotic large-strength magnetic regime $BM\gg1$. 

\bigskip
\noindent
{\bf ACKNOWLEDGMENTS}
\bigskip

This research is supported by the Carmel Science Foundation. I would
like to thank Yael Oren, Arbel M. Ongo, Ayelet B. Lata, and Alona B.
Tea for helpful discussions.



\begin{thebibliography}{99}
\bibitem{Notechar} The physically interesting case of central charged black holes with spatially regular horizons 
that support non-trivial scalar field configurations with a non-minimal coupling to the Maxwell electromagnetic invariant has 
been studied in \cite{Hersc1,Hersc2,Hodsc1,Hodsc2,Moh,Hodnrx}.
\bibitem{Noteetaa} Here $\eta$ is the non-minimal coupling parameter of the composed 
Einstein-Gauss-Bonnet-scalar field theory, see Eq. (\ref{Eq10}) below.
\bibitem{Notecord} Here $(t,r,\theta,\phi)$ are the familiar Schwarzschild spacetime coordinates.
\bibitem{Noteunits} We shall use natural units in which $G=c=1$.
\bibitem{NoteMev} Due to the asymptotic properties of the Melvin Universe \cite{Hernw,Melv,HF} 
it cannot describe the full spacetime of astrophysically realistic black holes. 
However, as nicely emphasized in \cite{Hernw}, the magnetized Schwarzschild-Melvin solution 
is expected to capture, approximately, the essential physical properties of the composed 
black-hole-scalar-field system in the vicinity of the black-hole horizon.
\bibitem{Noteett} The non-minimal coupling parameter $\eta$ has the dimensions of length$^2$.
\bibitem{Notesim} Here we have used the identities $\sin^2(2\theta)=4\sin^2\theta\cos^2\theta=4\sin^2\theta(1-\sin^2\theta)$.
\bibitem{Notedu} Note that we have used the relation (\ref{Eq18}) in Eq. (\ref{Eq19}).
\bibitem{Notering} Note that, due to the azimuthal symmetry of the magnetized Schwarzschild-Melvin black-hole spacetime 
[see the curved line element (\ref{Eq4})], the critical angular relation (\ref{Eq21}) describes two 
non-equatorial scalar rings with $\phi\in[0,2\pi]$.
\bibitem{Notesymm} The second classically allowed region for the near-horizon spontaneous scalarization phenomenon 
in the large-magnetic $BM\gg1$ regime is characterized by the narrow near-polar angular interval 
${{\sqrt{1-\sqrt{{{2}\over{3}}}}}\over{BM}}\leq\pi-\theta_{\text{scalar}}
\leq{\sqrt{{1+\sqrt{{{2}\over{3}}}}}\over{BM}}$.
\end{thebibliography}
\end{document}